\documentclass[bibauthoryear]{aa}

\usepackage{amssymb}
\usepackage{soul}

\newcommand{\epmax}{E_{p,{\rm max}}}
\newcommand{\barmax}{\bar{E}_{p,{\rm max}}}
\newcommand{\smax}{\sigma_{E p,{\rm max}}}

\usepackage{graphicx}
\usepackage[varg]{txfonts}
\usepackage{amsmath}
\usepackage{natbib}
\usepackage{xcolor}
\usepackage[colorlinks=true,citecolor=blue,linkcolor=blue,urlcolor=blue,anchorcolor=blue]{hyperref}

\begin{document}

	\title{What can cosmic-ray knees reveal about source populations? }
	
	\author{Myrto Falalaki
		\inst{1,2}
		\and
		Vasiliki Pavlidou\inst{1,2}
		}
	
	\institute{
		University of Crete, Department of Physics \& Institute of
		Theoretical \& Computational Physics, 70013 Herakleio, Greece \\
		\email{ph5690@edu.physics.uoc.gr; pavlidou@physics.uoc.gr}
		\and 
		Institute of Astrophysics,
		Foundation for Research and Technology-Hellas, 71110 Heraklion, Crete, Greece
		}
	
	\date{Received / accepted }

	\abstract
	{Breaks in the cosmic ray (CR) flux spectrum encode information on the properties of CR accelerator populations producing the observed flux. Spectral steepenings, known as knees, are  generally accompanied by a transition to heavier composition.}
{We seek generic features of CR source populations that imprint onto  knee observables robustly enough  to be discernible even in the presence of significant uncertainties in CR data. We explore how diversity among population members imprints on knee phenomenology, under the assumption that a knee is due to a fixed-rigidity cutoff in the source spectrum. Our scope is explicitly exclusionary: we do not fit specific datasets; rather, we ask which observed spectral features are incompatible with a single-population, fixed-rigidity cutoff picture, pointing toward additional physics.
}
{We use a simple theoretical model for a population of CR accelerators. Each member of the population accelerates CR stochastically to a power-law spectrum, up to a cutoff rigidity, resulting from source-confinement requirements. We allow for variance among members, in the cutoff rigidity and in the power-law slope. }
{We find that: (a) the slope step of the flux spectrum is $\sim 0.5$, decreasing weakly with increasing spread in either property; (b) composition always breaks first; (c) the difference between the break energies in composition and flux increases with increasing diversity. 
These trends are robust under our assumptions; deviations from them in observed data would indicate more complex physics than encoded in our simple model. }
{Comparing these trends with observed CR knees, we conclude that: (i) the primary knee at $\sim 4\times10^{15}$ eV is consistent with a constant-rigidity cutoff according to KASCADE-Grande data processed with post-LHC hadronic models, but not according to other datasets; (ii) the second knee at $\sim 5 \times 10^{17}$ eV conclusively requires more complexity than the cutoff of a single CR source population; (iii) a constant-rigidity source cutoff interpretation of the spectral feature identified by Auger at $\sim 10^{19}$ eV  cannot be rejected, provided there is a substantial spread in both cutoff rigidity and slope in the parent source population. Interestingly, a significant spread in slope would also result in spectral curvature before the break, which could in turn be contributing to the ankle feature.}
	

	\authorrunning{Falalaki \& Pavlidou}
	
	\maketitle
	
	
	\section{Introduction} \label{sec1}

    The energy spectrum of cosmic radiation is a critical observable for our understanding of the nature of cosmic ray (CR) accelerators.  This spectrum can  be described as a power law over many orders of magnitude (from $\sim10^9$ eV to $\sim10^{20}$ eV). The spectral power index, however, exhibits changes at characteristic energies, referred to as knees (when the spectrum steepens) and ankles (when the spectrum flattens).

 Knees  have now been spectrally resolved in great detail at several different energies in the CR spectrum, ranging from few times $10^{15}$ eV to $10^{19}$ eV (e.g., \citealp{Akeno, Flyseye1990, CASABLANCA2001, Hiresmia2001, KASCADE2005, HiresStereo, TibetIII,  IceTop,   AugerNewFeature, LHAASO2024}). These breaks are typically accompanied by transitions to heavier composition around the same energies as the spectral break (e.g., \citealp{KASCADE2019, AugerNewFeature}). This behavior has led to the qualitative interpretation of the knee phenomenon as a constant-rigidity cutoff (also referred to as Peters cycle, \citealp{Peters}), either in Galactic confinement or in source acceleration / confinement: particles accelerated stochastically by a CR source cannot reach energies beyond the threshold where magnetic confinement is lost . This effect occurs roughly at the energy where a particle's gyroradius in the source's magnetic field becomes comparable to the size of the source (e.g., \citealp{Hillas1984}). 
 
A rich literature exists on quantitative models of the various knee-like breaks of the CR spectrum, ranging from fits of specific datasets with complex phenomenological models, to comprehensive physical models, including specific hypotheses on source accelerator physics and source population properties, propagation effects, and simulations of atmospheric air showers to obtain direct observables (e.g., \citealp{HillasNotHillas,VarietyOfSupernova2003, Jorg2004, Lemoine2005, KumikoLemoine2008, ProgenitorModel2016, CombinedFitOriginal, Kohta2018, SilviaEsteban, CombinedFitUpdate2022, BlandfordGlobus23, LuisUnger24}). 

Despite these intensive theoretical and experimental efforts, the exact characteristics, nature and origin of {\em all} observed CR knees is still under debate. One very important factor contributing to this continued uncertainty is that the data are not yet fully converged. High-level observables (i.e. spectra and moments of the log mass-number distribution) can differ between observatories using different observational techniques at the same energies (see e.g. review of data on the primary knee in \citealp{BlumerReview2009}); between different observatories using similar techniques (see, e.g., differences in flux and composition at the highest energies reported by the Pierre Auger Observatory and Telescope Array, \citealp{AugerTASpectrum, AugerTAComposition}); between events recorded by the same observatory when using different observables (see, e.g., differences in composition at ultrahigh energies derived when using shower depth or shower muon content, e.g., \citealp{MuonXmax}); and even between identical recorded datasets when processed with different simulations of hadronic interactions (for example, with pre- versus post-LHC models, \citealp{KASCADE2005, KASCADE2019}; or with different hadronic interaction packages of the same generation, \citealp{Xmax_lnA}). As a result, detailed fits to sophisticated source population models may fail or return parameters that appear astrophysically contrived (e.g., \citealp{CombinedFitOriginal, FoteiniUnger}), not because the models are necessarily inconsistent with the actual source populations, but rather because our data from said source populations might be affected by systematics that are not adequately quantified or accounted for. 
 
The question then arises whether there exist any generic features of cosmic-ray source populations that imprint onto cosmic-ray observables in a manner that is robust enough and distinct enough to be discernible even in such a still-fluid experimental landscape. This is our aim in this paper. Using the simplest possible model for an underlying population of CR accelerators, we seek to build insight on how  the diversity between population members could imprint on the knee phenomenology, under the assumption that a knee is a fixed-rigidity-cutoff phenomenon. Such insight is necessary in order to identify any robust features and trends of simple population models and their most straightforward variations.  Absence of said features would then constitute evidence of more complex physics of either CR acceleration (complicated astrophysics) or CR interactions at the detection site (unexpected particle physics). Throughout, we therefore use a deliberately minimal model to test falsifiability: if a feature violates the robust trends of this simple picture, the population-effects-only hypothesis can be rejected, even though consistency never proves sufficiency. We will refer to such use (ruling out a single-population fixed-rigidity picture based on the absence of characteristic expected spectral and composition features) as an exclusion test.

 A knee is described phenomenologically by the characteristic energy where the break occurs in the CR flux spectrum, by the spectral slopes before and after this characteristic energy, and the (generally different) energy where the accompanying break in CR composition takes place.  We will investigate whether features of the population of CR sources responsible for the knee result in patterns in these observables that are simple and robust enough that they might be recognizable even in the presence of significant systematic uncertainties in the data. 
 
 To this end, we set up a simple, generic model for a population of CR accelerators with the following properties: (a) For energies well below its rigidity cutoff (equal to the proton cutoff energy,  $\epmax$), each source contributes to the Galactic CR flux particles with a power-law spectrum of energies of slope $\gamma$ (encoding both the source acceleration properties, and losses during propagation). (b) The rigidity cutoff can be described by an exponential suppression of the power-law spectrum. Different nuclei cutoff at different energies scaling as $Z\epmax$. (c) We allow diversity between members of the CR source population in both $\epmax$ and $\gamma$. 
A "knee" observed in the summed CR spectrum due to such a population of sources will consist of: a break in the spectrum, encoding, but not necessarily equal to, $\epmax$; and a break in the composition at a similar, but not necessarily identical, energy. The location in energy of those breaks, and the slope difference between and after the knee will be modulated by the distribution of $\gamma$ and $\epmax$ among the individual members of the population. 

We explore the features and trends of this simple model, and in particular we address the following specific questions: (1) Under what conditions do the flux spectrum and composition spectrum break together? (2) When they do not, which one breaks first? (3) How do the break energies relate to $\epmax$? (4) How does the diversity of the population properties, manifesting as a spread in $\epmax$ and $\gamma$, affect the answers to these questions?

Our model intentionally omits additional processes (e.g., energy losses and interaction channels at the highest energies, multi-population transitions, or detailed transport changes), because we aim to isolate and demonstrate the generic imprint of source-population diversity alone. The result we obtain in this way  is a set of robust qualitative trends and their energy ordering; their absence in data is a diagnostic of missing physics in this minimalist picture.

The simplicity of our model is by design: we seek features imprinted by these physical phenomena alone, and so we attempt to isolate their effect by stripping the model we use down to basics. Our framework is deliberately single-population and omits multi-population transitions (e.g. Galactic-to-extragalactic extragalactic or multiple Galactic components); when such transitions occur, the qualitative trends we derive are intended as exclusion tests of the single-population hypothesis rather than as a fit to that regime.

This paper is organized as follows. In \S \ref{sec2} we lay out the formulation of our model, emphasizing its free parameters and their impact on the observable flux and composition spectra. In \S \ref{sec3}
we explore the behavior of these observables as the  population diversity increases. We summarize our conclusions and discuss them in the context of observations of various CR knees in \S \ref{sec4}.

\section{The model} \label{sec2}
\subsection{Single Source}
	We implement a fixed-rigidity cutoff, characteristic of CR sources relying on magnetic confinement
 \citep{Hillas1984, Jorg2004}.
 We do so by assuming that  the differential flux $F_z$ produced by a single cosmic ray source as a function of the energy E of primary particles of charge $Z$ has the form:
    \begin{equation}
	   F_{z}(E) = F_{0}(Z)\left(\dfrac{E}{E_0}\right)^{-\gamma} \exp\left[ -\dfrac{E}{Z \epmax }\right]
	   \label{5}
    \end{equation}
    where  $\epmax$ is the rigidity cutoff, $F_0(Z)$ is the flux of a specific species of  atomic number Z at some normalization energy $E_0 \ll \epmax$,  and $\gamma$ is the low-energy power-law slope. In our model, we assume that $\gamma$ is identical for all species in a single source, and that it encodes both acceleration and propagation physics (including losses and escape). The validity of this approach is established observationally, as, for example, the cosmic ray spectrum sufficiently below the primary knee can be well described by a single power law, with slope encoding both acceleration and energy-dependent propagation losses and escape. We emphasise, however, that in our context, $\gamma$ is meant to  encode physics at energies right below each knee, before the maximum rigidity is reached for any species. While in our plots in later sections we extend the depiction of the spectrum behaviour over several orders of magnitude below the knee, we do so solely to improve visualization of the break feature.

The total flux from a single source at an energy $E$ will be the sum of $F_z(E)$ over all primary charges:
    \begin{equation}
	   F(E) =\sum_{Z}F_{z}(E)\,.
	   \label{6}
    \end{equation}
In this work, we have taken the relative abundances $F_0(Z)$ from \cite{Zhao2015} (see their table 1, where they present their fitted parameters for a power-law like model for every nuclei). While \cite{Zhao2015} fit a different slope $\gamma_z$ for different species, we have adopted here an effective value of $\gamma\approx 2.66$ for all species. However, our results are independent of the exact choice for the value of $\gamma$, since we always plot deviations (differences) of fitted slopes from the underlying source $\gamma$.  We simulate fluxes for the following individual species: H, He, C, O, Ne, Mg, Si and Fe. These abundances are observationally motivated for energies below the primary CR knee; however our qualitative conclusions do not depend sensitively on this choice, and so they hold for any source population accelerating particles of mixed composition roughly comparable to that of Galactic CR, even if the pre-break abundances differ in their details from the ones we have adopted here. To verify that our qualitative conclusions are not contingent on the adopted pre-cutoff composition, we repeat the analysis with abundances tuned to higher energies (KASCADE-Grande; \citealp{New_composition}).

We quantify the resulting composition spectrum by the average of the logarithm of the mass number as a function of energy, $\langle \ln A\rangle$ (E), since this metric is frequently used to summarize CR composition observations. For a single source, this will be given by
  \begin{equation}
\langle\ln{A}\rangle (E) = \frac{\sum_{Z} F_{Z}(E) \ln (A_Z)}{\sum_{Z} F_{Z}(E)}\,.
        \label{comp1}
    \end{equation}

\subsection{Source population} \label{sec2.1}

The CR observables on Earth at a given energy are produced by particles accelerated by a population of sources with a distribution of sizes, B-fields and other properties. Variations in these  physical conditions among population members will result in corresponding variations of source model parameters \citep[see, e.g.,][]{Workman2022,Diesing2023}. In our simple source model, these parameters are $F_0(Z)$, $\gamma$, and $\epmax$. In this work, we are interested in any signatures of 
the diversity in $\epmax$ and $\gamma$ imprinted on the observables of a  knee produced by a constant-rigidity cutoff in the spectra of sources. To isolate the effects of each of these source properties, we explore the effect of increasing spread in one of $\epmax$, $\gamma$, while keeping the distribution of the other fixed to a delta function. We also make the simplifying assumption that  relative values of $\tilde{f}_0(Z)=F_0(Z)/F_0(Z=1)$ (the accelerated particles relative abundances at energies well below the proton cutoff) are identical among different population members, so that the only quantities that may vary between individual CR sources are $F_0(Z=1)$,  $\epmax$, and $\gamma$.  

Mathematically, we model this picture as follows. Choosing a normalization energy $E_0$ well below the lowest $\epmax$ encountered in the specific population, the exponential suppression factor in Eq.~(\ref{5}) at $E_0$ is equal to $1$ for all sources. The resulting total flux from all population sources at that energy 
can be calculated as:
\begin{equation}
   F_{Z,tot}(E_0\ll \min{\epmax}) =\sum_{i} F_{0,i}(Z=1) 
   \tilde{f}_0(Z) \equiv F_{Z,tot,0}\,,
   \label{totnormdef}
\end{equation}
where the summation is over different population member sources. This equation defines the flux normalization of species $Z$ for the population, $F_{Z,tot,0}$. 

To calculate the spectrum at higher energies, we introduce
the probability distributions  $p_E(\epmax)$ and $p_\gamma(\gamma)$ of the cutoff rigidity, and the low-energy power-law slope, respectively, in the population. Formally, $p_E(\epmax)d\epmax$ is the fraction of particles at energy $E_0$ that were accelerated by sources with rigidity cutoffs between $\epmax$ and $\epmax+d\epmax$; and $p_\gamma(\gamma)d\gamma$ is the fraction of particles accelerated by sources that, had they been responsible for the entire CR spectrum at low energies, would have produced  (after acceleration and propagation) a  spectrum which would have been a superposition of power laws with slopes between $\gamma$ and $\gamma + d\gamma$. 
Then, the total flux of species $Z$ due to the entire population can be calculated through 
    \begin{equation}
       F_{Z, {\rm tot}}(E)\!
        = \!F_{Z, tot,0} \!\!\!
        \int_{0}^{\infty}\!\!\!\!\!\! d\epmax p_E(\epmax)
        {\rm e}^{-\dfrac{E}{Z\epmax}} \!\!\!
        \int_{0}^{\infty}
        \!\!\!\!\! d\gamma p_{\gamma}(\gamma)
 \left(\frac{E}{E_0}\right)^{-\gamma} 
 \!\!.
 \label{doubleintegral}
    \end{equation}
The total CR flux due to all species  will be 
\begin{equation}
    F_{\rm tot}(E)
    = \sum_Z     F_{z, {\rm tot}}(E)\,.
        \label{fluxpop}
\end{equation}
Finally, the composition spectrum due to the population will be given by 
  \begin{equation}
\langle\ln{A}\rangle_{\rm tot} (E) = \frac{\sum_{Z} F_{Z, {\rm tot}}(E) \ln (A_Z)}{\sum_{Z} F_{Z, {\rm tot}}(E)}\,.
        \label{comp2}
    \end{equation}
    
 \subsection{Varying $\epmax$}
 We will estimate the effect of a spread in $\epmax$ among population members by assuming $p_E(\epmax)$ is lognormal: 
    \begin{equation}
        p_E(\epmax) = \dfrac{1}{\epmax\sigma \sqrt{2\pi}} \exp \left[ -\dfrac{(\ln \epmax-\mu)^2}{2\sigma^2}\right]\,.
        \label{9}
    \end{equation}
The parameters of the distribution $\mu$ and $\sigma$ are related to the mean and the standard deviation of the distribution through 
   \begin{equation}
        \mu = \ln \left(\frac{\barmax ^{2}}{\sqrt{\barmax^2+\smax^{2}}}\right)
        \label{10}
    \end{equation}
    and 
    \begin{equation}
        \sigma^2 = \ln \left[1+\left(\dfrac{\smax} {\barmax}
        \right)^2\right]\,.
       \label{11}
    \end{equation}
The advantage of the lognormal distribution is that $\epmax$ is positive definite and as a result arbitrarily large values of $\smax$ can be accommodated, allowing us to examine asymptotic behaviors for very diverse populations. The distribution $p_{\gamma}$ remains fixed to a delta function.

\subsection{Varying $\gamma$}
To estimate the effect of a spread in $\gamma$, we keep $p_E(\epmax)$ fixed to a delta function, 
and we implement 
a Gaussian  $p_\gamma(\gamma)$, with mean $\bar{\gamma}$ and spread $\sigma_\gamma$. In this case, Eq.\eqref{doubleintegral} has an analytic solution,
     \begin{equation}
        F_{Z, {\rm tot}}(E)
        =  
F_{z,tot,0} \exp \left[ -\frac{E}{Z\epmax}\right]
 \left(\frac{E}{E_0}\right)^{-\bar{\gamma}}\!\!\exp 
 \left[ -\frac{\sigma^{2}_\gamma}{2}\ln^2\left( \dfrac{E}{E_0}\right) \right].
    \end{equation}

\subsection{Observables}

Even though the flux suppression of individual CR species is exponential, the fact that the suppression sets in at increasing energy for increasing $Z$  results to a flattening of the knee. This will be true for individual sources (Eq.~\ref{6}), and more so for a population (Eq.~\ref{fluxpop}) where the spread in $\epmax$ and/or $\gamma$ will result in a smoother break. Observed knees can be fitted well by broken power laws, and we attempt the same for our model results. In particular, we test whether 
a broken power law of the form:
    \begin{equation}
        F_{tot}(E) = C \times
        \begin{cases}
        \left(\dfrac{E}{E_b}\right)^{-\gamma_1}\,, & E \leq E_b \\
        \left(\dfrac{E}{E_b}\right)^{-\gamma_2}\,, & E > E_b
        \end{cases}
        \label{12}
    \end{equation}
    can describe adequately the total flux around the knee. From this fit, we exctract "observables" $\gamma_1$ (slope before the break), $\gamma_2$ (slope after the break), and $E_b$ (break energy). 

    We additionally evaluate a fourth "observable", the composition break energy, $E_A$. We do so by:
    (a) fitting the composition spectrum of Eq.~(\ref{comp2}) for $E\ll \barmax$ by a constant (horizontal line); and
    (b) fitting the composition spectrum  by a logarithmic increase (linear increase with $\log E$) for  $E\gg \barmax$. Then, $E_A$ is the energy at which the two lines intersect (see lower-right panel of Fig.~\ref{fig:fig2}). 
    
	\section{Results} \label{sec3}

We start by examining the trends induced on the "observables" by a gradually increasing spread in $\epmax$ among population members.

  \begin{figure*}[h]
        \centering
        \includegraphics[width=\textwidth]{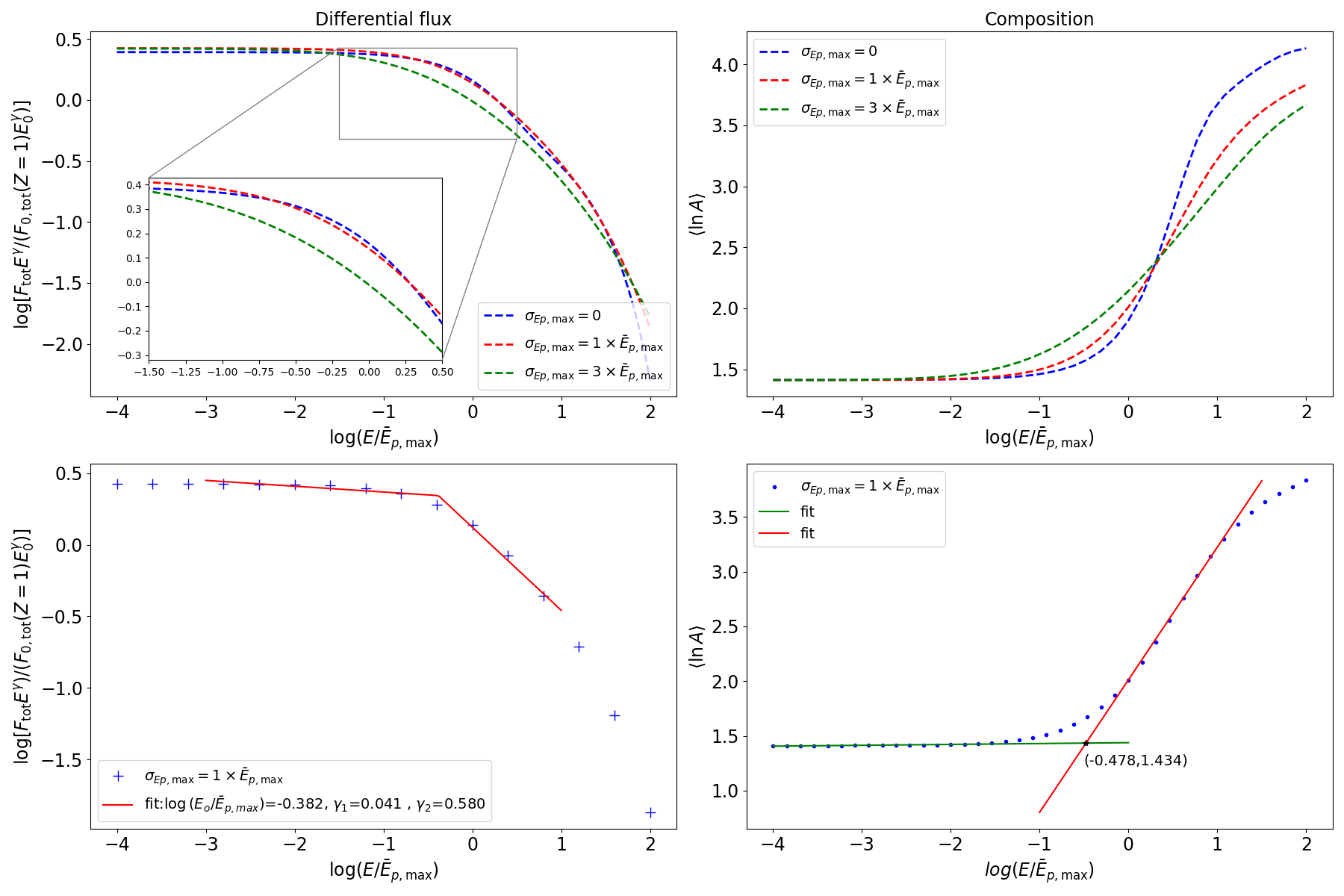}
        \caption{All-particle flux and composition spectra 
        in the case of a lognormal $p(\epmax)$. Upper-left panel: all-particle flux  spectrum flattened by $E^{\gamma}$. Upper-right panel: composition spectrum. Lower-left panel: broken power-law fit around the knee for the flux spectrum. Crosses correspond to mock observations obtained from our model. The solid line is the fit of Eq.~(\ref{12}). Lower-right panel: low- and high-energy asymptotic logarithmic fits to the composition spectrum (green and red solid lines respectively). Points again correspond to mock data obtained from our model. The composition break energy given by the intersection of the two lines (black star).  }
        \label{fig:fig2}
    \end{figure*}
 
  In the upper left panel of Fig.~\ref{fig:fig2} we show the all-particle spectrum (flattened by $E^{\gamma}$) as we evaluate it from Eqs.~\eqref{doubleintegral} and \eqref{fluxpop} and for a lognormal $p(\epmax)$, for $\smax \in[0,3\barmax]$. For broader $p(\epmax)$, the all-particle spectrum deviates sooner (at lower energies) from its low-energy asymptotic behavior, as a result of the property of the lognormal distribution to peak around values of $\epmax < \barmax$. At the same time, the spectrum falls less steeply at high energies, owing to the tail of the lognormal towards high $\epmax$ values.

\begin{figure}[h]
        \centering
        \includegraphics[width = 0.5\textwidth]{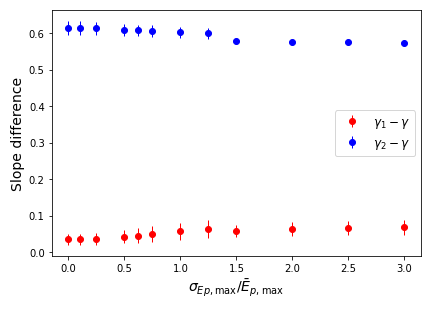}
        \caption{Values of the spectral power indices $\gamma_1$ and $\gamma_2$, indicated with red and blue points respectively, as a function of $\smax/\barmax$.}
        \label{fig:fig3}
    \end{figure}

\begin{figure}[t!]
        \centering
        \includegraphics[width = 0.5\textwidth]{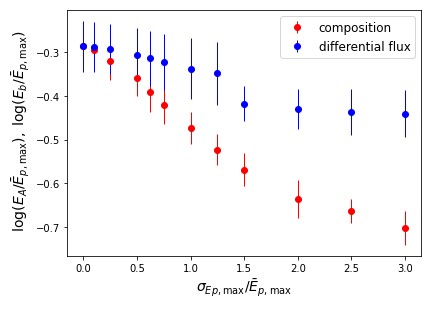}
        \caption{Energies of the composition break (red) and the flux break (blue) in units of $\barmax$ as a function of $\smax/\barmax$.}
        \label{fig:fig4}
    \end{figure}

In the lower-left panel of Fig.~\ref{fig:fig2}, we show an example of fitting the all-particle flux spectrum around the knee with a broken power law.  The case depicted here is produced by setting $\smax =\epmax$ in the lognormal $p(\epmax)$.  It is through such fits that we obtain the break point $E_b$ and the spectral power indices $\gamma_1 $,$\gamma_2$ (the "observables" discussed in the context of Eq.~\ref{12}), which we present and discuss in Figs. 
\ref{fig:fig3} and \ref{fig:fig4} as functions of $\smax$.

The upper-right panel of Fig.~\ref{fig:fig2} shows the composition spectrum for the same models as in the upper-left panel. The break in flux is accompanied by a break in composition. The impact of increasing $\smax$ on the composition spectrum is more pronounced than that on the flux spectrum, both towards lower and higher energies, as well as in terms of the steepness of the break. To quantify this behavior, we again fit the low-energy and high-energy trends with logarithmic functions ($\langle \ln A \rangle$ linear in $\log E$), as shown in the lower-right panel of Fig.~\ref{fig:fig2}. The intersection of the two fits defines the composition break energy, $E_A$.

The effect of $\smax$ on the "observables" is shown in Figs.~\ref{fig:fig3} and \ref{fig:fig4}. 
Figure \ref{fig:fig3} shows the exponents $\gamma_1$ and $\gamma_2$ (slopes before and after the break, red and blue points respectively), evaluated as difference from the single-source slope $\gamma$.  The error budget is completely dominated by systematic uncertainties, primarily driven by the choice of points to include in the fit. Here, error bars  correspond to the difference between including or dropping an extra point at the high-energy tail of the fit (see lower-left panel of Fig.~\ref{fig:fig2}).
 
The difference between the two slopes starts at $\sim 0.6$ for a population comprised of members with identical $\epmax$ ($\smax =0$), and decreases slowly as $\smax$ increases, reaching $\sim 0.5$ for $\smax = 3\barmax$. This mild trend is a result of two compounding effects. First, as $\smax$ increases, the number of sources with low $\epmax$ also increases, so the low-energy branch of the broken power law is also affected (becomes steeper, $\gamma_1$ increases) as some sources have already started becoming suppressed at low energies. Second, a high $\smax$ also results in a larger number of sources with high $\epmax$. The flux of these sources is not suppressed until higher energies, resulting in a smoother decline of the all-particle spectrum. The high-energy branch of the power law thus becomes shallower ($\gamma_2$ decreases). The overall conclusion is that a fixed-rigidity cutoff in a CR source population with identical power-law slopes $\gamma$ and a pre-break composition roughly comparable to that of Galactic CR produces a knee with a slope change in the range of $\sim 0.5-0.6$, regardless of spread in $\epmax$. 

In Fig.~\ref{fig:fig4} we overplot the break energies of the spectrum ($E_b$, blue points) and of the composition ($E_A$, red points) as a function of the spread in $\epmax$. Error bars are again dominated by systematics. For $E_b$, they are obtained in the same way as the error bars in $\gamma_1$ and $\gamma_2$. In the case of $E_A$, they correspond to the difference resulting from retaining or dropping one point towards the break in the rising part of the composition spectrum. We observe that in the case of a population with members of identical, or very similar, $\epmax$, composition and flux spectrum break together, at an energy about half of $\epmax$. As $\smax$ increases both the composition and the flux spectrum break earlier, however $E_A$ decreases faster than $E_b$, and as a result {\em composition always breaks first}. 

    \begin{figure}[h]
        \centering
        \includegraphics[width = 0.5\textwidth]{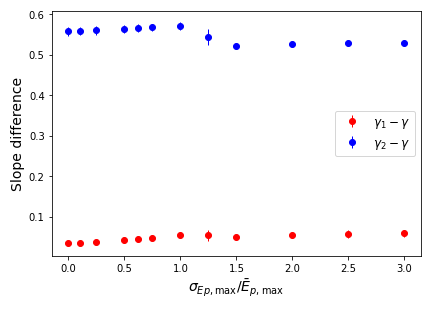}
        \caption{As in Fig.~\ref{fig:fig3}, but using initial CR composition taken from \citet{New_composition}, which is more appropriate for energies above the knee. 
        \label{gammas_new_comp}}
    \end{figure}

    \begin{figure}[h]
        \centering
        \includegraphics[width = 0.5\textwidth]{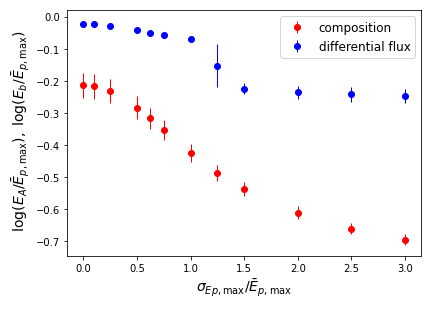}
        \caption{As in Fig.~\ref{fig:fig4}, but using initial CR composition taken from \citet{New_composition}, which is more appropriate for energies above the knee. }
        \label{breaks_new_comp}
    \end{figure}

 To examine whether the trends we identify depend strongly on our assumed elemental composition, we repeated the analysis using an alternative set of relative abundances derived from \citealp{New_composition}, and following exactly the same procedure as described in Section \ref{sec2}.
The new versions of Fig. \ref{fig:fig3} and \ref{fig:fig4} are presented in Fig. \ref{gammas_new_comp} and \ref{breaks_new_comp}, respectively. 
The qualitative behavior of all observables remains unchanged: the slope step stays within the same range ($\sim0.5 - 0.6$), and composition continues to break before the flux. Therefore, our main conclusions are robust with respect to reasonable variations in the assumed source abundances.

 \begin{figure}[h]
        \centering
        \includegraphics[width = 0.5\textwidth]{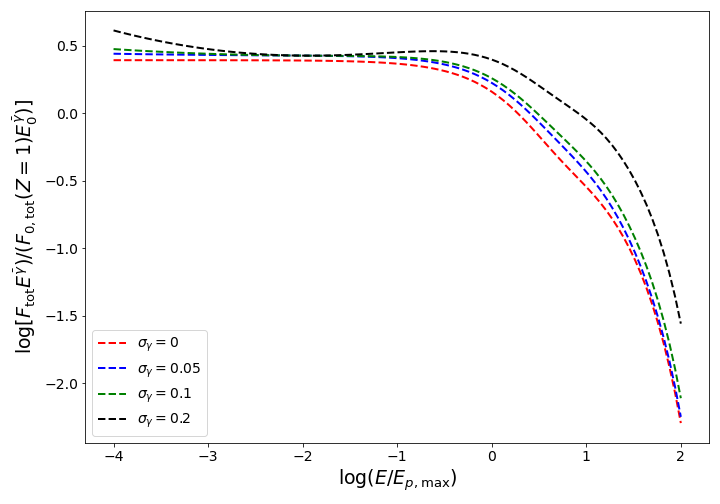}        \caption{All-particle spectrum flattened by $E^{\bar{\gamma}}$, as produced by Eq.~(\ref{fluxpop})  with a Gaussian $p_\gamma(\gamma)$ and a delta-function $p_E(\epmax)$. Different lines correspond to different values of $\sigma_\gamma$, shown in the legend. \label{fig_gamma_lines}}
    \end{figure}

        \begin{figure}[h]
        \centering
        \includegraphics[width = 0.5\textwidth]{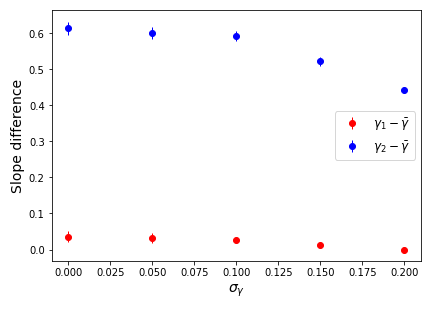}
        \caption{Values of the spectral power indices $\gamma_1$ and $\gamma_2$, indicated with red and blue points respectively, as a function of $\sigma_\gamma$. \label{gamma_gamma}}
    \end{figure}

        \begin{figure}[h]
        \centering
        \includegraphics[width = 0.5\textwidth]{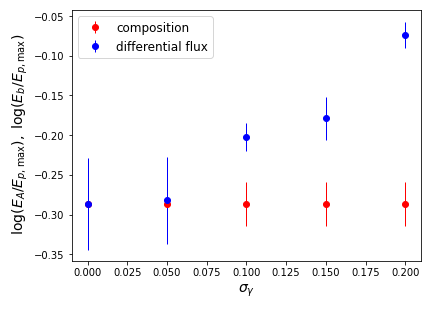}
        \caption{Energies of the composition break (red) and the flux break (in blue) in units of $\epmax$ as a function of $\sigma_\gamma$.\label{gamma_breaks}}
    \end{figure}

We now turn to trends in "observables" that result from a spread in $\gamma$ among population members. For these calculations, we keep $p_E(\epmax)$ fixed to a delta function. In Fig.~ \ref{fig_gamma_lines} we plot the all-particle spectrum (flattened by $E^{\bar{\gamma}}$) as we evaluate it from Eqs.~\eqref{doubleintegral} and \eqref{fluxpop} and for a Gaussian  $p_\gamma(\gamma)$ .  Different line colors correspond to different values of $\sigma_\gamma$. Here, the range of $\sigma_\gamma$ we consider is much narrower than the range of $\smax$. The reason is that a large spread in power law indices results in significant spectral curvature (see e.g. black dashed line in Fig.~\ref{fig_gamma_lines}), which is not generally seen in the CR spectrum. Interestingly, small spreads in $\gamma$ appear to generate  diversity in the flux spectrum comparable to that produced by very substantial spreads in $\epmax$. 

We investigate the qualitative direction of the  trends seen in the "observables" with increasing $\sigma_\gamma$, and quantify them, in Figs.~\ref{gamma_gamma} and \ref{gamma_breaks}. The trend of the sharpness of the break with $\sigma_\gamma$ is shown in  Fig.~\ref{gamma_gamma}. Colors and error bars are as in Fig.~\ref{fig:fig3}. Here again the slope step decreases with increasing population spread, from about 0.6 for a population with no spread, to about 0.45 for $\sigma_\gamma=0.2$, driven primarily by the post-break slope becoming shallower. Still, the effect is very mild: in a standard fixed-rigidity-cutoff knee, the slope change does not become very different from a 0.5 step, even if the underlying accelerator population exhibits significant spread in its properties. 

The trends in energy breaks however are now different, as is shown in Fig.~\ref{gamma_breaks}. 
The location of the composition break, $E_A$, is unaffected by any spread in $\gamma$. 
The flux spectrum break, $E_b$, on the other hand {\em increases} with increasing population spread: the more diverse in $\gamma$ the population, the later the flux spectrum breaks. Importantly, the overall result goes in the same direction as in the case of a spread in $\epmax$: {\em composition always breaks first.}

    \section{Conclusions and Discussion } \label{sec4}

 We have used the simplest possible model of a knee induced by a fixed-rigidity CR-source cutoff, in order to explore how diversity among the members of the underlying cosmic-ray accelerator population affects the knee phenomenology. In particular, we explored how diversity in (a) the rigidity cutoff $\epmax$ and (b) in the pre-knee cosmic-ray slope $\gamma$ (encoding both  acceleration and loss/propagation physics) is imprinted in (i) the difference in pre- and post-break flux spectrum slopes, and (ii) in the break energies of the flux and composition spectra. We have identified the following robust trends.\\
 1. A knee induced by a fixed-rigidity cutoff in the CR source population exhibits a steep break in the all-particle spectrum, with a slope step around 0.5. Diversity in the population (in either $\epmax$ or $\gamma$) tends to somewhat reduce the steepness of the break, but the effect is weak. \\  
 2. Diversity in $\epmax$ moves both the flux break energy, $E_b$, and the composition break energy, $E_A$, to lower values, with the composition being more strongly affected. In contrast, diversity in $\gamma$ leaves $E_A$ practically unaffected, but moves $E_b$ to higher energies.\\
 3. Any diversity in either $\epmax$ or $\gamma$ works to separate $E_b$ from $E_A$, in the same direction: {\em composition breaks first}. The difference between the two can be as large as a factor of several, especially if both $\epmax$ and $\gamma$ vary substantially among population members.  \\
 Any qualitative deviation from these trends would require a model of significantly higher astrophysical complexity than the one we discussed here.
 
  There exist several energy regimes where such complexity is not only possible, but likely. Above $\sim 10^{19}$ eV, photo-hadronic and photo-disintegration interactions for protons and nuclei (GZK suppression, \citealp{G,ZK}) shape the spectrum and composition independently of population effects. Pair-production losses above $10^{18}$ eV can produce an ankle-like feature, depending on composition (e.g., \citealp{Berezinsky2006}). Transitions between distinct source classes (e.g., Galactic vs extragalactic, or between distinct extragalactic popualtions, e.g., \citealp{CombinedFitOriginal}) can also impart features in CR spectrum and composition. Our framework does not model these channels; rather, it is intended to test whether population effects alone could be the dominant driver of a given feature. When high-energy processes co-act with population diversity, the qualitative trends we identify still provide exclusion tests. For example, a composition break preceding a spectral steepening below the GZK cutoff is a natural outcome of a population cutoff and remains consistent even if a separate GZK suppression operates at higher energy.
 
 We can use these insights to discuss qualitatively the likelihood of each of the steepening features in the broadband cosmic-ray spectrum being a simple fixed-rigidity CR-source-cutoff knee, even without any detailed fits to a specific model or the underlying accelerator populations.Indeed, as the primary aim of this work has been to search for robust features imprinted on the spectrum as a result of  the specific physical processes that we included as ingredients in our model, and given the intentional simplicity of our assumptions, our final product and prediction are the spectral features and their energy ordering themselves, rather than the detailed form of the spectrum. In this sense, the meaningful comparison with present (or future) cosmic ray data is through exactly the comparison of our predictions with the presence and energy ordering of breaks in the data - which we perform below. 

For the primary CR "knee" at $\sim 4\times 10^{15}$ eV \citep{CASABLANCA2001, KASCADE2005, IceTop, LHAASO2024}, the situation is still unclear, as observations from different experiments and using different techniques have unfortunately not yet fully converged, even at the very coarse level needed for the type of comparison we are seeking to make in this work. For example, early results presented in the \citet{BlumerReview2009} review, but also very recent results from the {\it Large High Altitude Air Shower Observatory} (LHAASO) experiment \citep{LHAASO2024}, indicate a small slope step (between $0.2$ and $0.4$), and the composition breaking {\em after} the spectrum. In the context of a 
simple fixed-rigidity source-cutoff knee described here, a slope step on the low side could be conceivably achieved with a combination of variations in both $\epmax$ and $\gamma$ in the population of contributing sources. However, such a physical picture cannot accommodate the composition breaking after the spectrum.  Even if we were to accept that, within uncertainties, composition and spectrum could be breaking together, this would point towards a population with very little spread in both $\gamma$ and $\epmax$, which would then result to a sharper slope step than the one observed. This combination of phenomenological observables then hint towards for a more complex picture - plausibly, for example, towards more than one source populations contributing cosmic rays around the primary knee, or additional physical effects (e.g., \citealp{HillasNotHillas}).  On the other hand, in the latest, post-LHC, reanalysis of data from the KASCADE-Grande experiment \citep{KASCADE2017, KASCADE2019}, the composition appears to be already getting heavier before the break in spectrum, while the slope step is reported at $\sim 0.5$, as might be expected from the simplest version of fixed-rigidity knee explored here.  

For the "second knee" around $5\times10^{17}$ eV (see e.g. \citealp{BergmanReview} for a review), the observational situation is similarly unconverged.  Different datasets disagree over the location of the composition break, and on whether, before the break, the composition was getting heavier \citep{Flyseye1990} or lighter \citep{Hiresmia2001,HiresStereo}.  However there is consensus that the break is soft (slope step between $0.2$ and $0.3$), while the composition across the second knee is becoming {\em lighter}. Most likely then in this case there is a second, light (i.e., still efficiently accelerating) population contributing (e.g., \citealp{Heino}), so the simple physics we explored is not adequate to model this transition. 
 More specifically,  in ``dip/transition'' scenarios, the second knee marks the onset of the Galactic-to-extragalactic transition: the Galactic heavy component fades while a lighter extragalactic component appears; in that case, the ankle reflects pair-production shaping of the extragalactic proton spectrum. Alternative models explore two Galactic components below the ankle, with an extragalactic contribution becoming dominant only closer to $10^{18.5}$--$10^{19}$\,eV. Either way, a changing population readily dilutes the slope step to $\Delta\gamma\sim 0.2$--$0.3$ and can produce a lightening composition across the steepening---features incompatible with a single-population fixed-rigidity cutoff and therefore fully consistent with our exclusion result for the second knee \citep{Aloisio2008,Lemoine2005,Thoudam2016,Apel2011,Apel2013}.

The knee-like "new feature" identified by the Pierre Auger Observatory in the cosmic ray spectrum around $10^{19}$ eV \citep{AugerNewFeature} features a slope step of $\sim 0.5$, right in the expected range for a fixed-rigidity source-cutoff knee.  In this reading, the population-cutoff steepening would be distinct from, and preceding, the higher-energy GZK suppression, which would further steepen the spectrum at yet higher energies.
This scenario would further support the interpretation that the break seen in composition-sensitive observables {\em before} the break in the spectrum \citep{AugerNFComposition} indeed indicates a transition to heavier primaries, rather than a signature of exotic physics (as, in, e.g. \citealp{FarrarAllen} or \citealp{pavlidou_tomaras_2019}). This is further reinforced by the reconstructed shape of the $\langle \ln A \rangle$ spectrum. Although the overall normalization of the composition spectrum is very dependent on the choice for hadronic interaction modelling, the relative change of $\langle \ln A \rangle$ with energy does feature the overall shape seen in the upper-right panel of Fig.~\ref{fig:fig2}: $\langle \ln A \rangle $ flattens off between $10^{18}$ and $10^{18.5}$ eV, before starting to increase logarithmically around $10^{18.6}$ eV. Our interpretation is then in overall agreement with the one proposed by \citet{AugerNewFeature}. Interestingly, however, the difference between the locations of the composition break (around $10^{18.6}$ eV if we follow the same procedure we have used here to determine $E_A$) and the spectrum break (reported by \citealp{AugerNewFeature} at $10^{19.1}$ eV) is higher than the differences we have seen produced by variations in $\epmax$ alone. The implication is that $\gamma$ may also be significantly varying in the dominant CR source population at these energies. This in turn could produce a non-negligible curvature in the spectrum before the break (see black dashed line in Fig.~\ref{fig_gamma_lines}), that may be contributing to the curvature of the ankle, which is also located at the same energy as the composition break. 
	
	\begin{acknowledgements}
 
MF acknowledges support 
by the European Research Council under the European Union's Horizon 2020 research and innovation programme, grant agreement No. 771282 (PASIPHAE). VP acknowledges support by the Hellenic Foundation for Research and Innovation under the “First Call for H.F.R.I. Research Projects to support Faculty members and Researchers and the procurement of high-cost research equipment grant”,  Project 1552 CIRCE, and by the  Foundation of Research and Technology - Hellas Synergy Grants Program (project MagMASim). MF would like to thank Nick Loudas and Chryssi Koukouraki for their insightful comments and discussions. 

	\end{acknowledgements}

	\bibliography{references}

\end{document}